# Incoherent digital holograms acquired by interferenceless coded aperture correlation holography system without refractive lenses

Manoj Kumar\*, A. Vijayakumar, and Joseph Rosen

[1]Department of Electrical and Computer Engineering, Ben-Gurion University of the Negev, P.O. Box 653, Beer-Sheva 8410501, Israel.
\*Corresponding author: manojklakra@gmail.com

**ABSTRACT**

We present a lensless, interferenceless incoherent digital holography technique based on the principle of coded aperture correlation holography. The acquired digital hologram by this technique contains a three-dimensional image of some observed scene. Light diffracted by a point object is modulated using a random-like coded phase mask (CPM) and the intensity pattern is recorded and composed as a point spread hologram (PSH). A library of PSH is created using the same CPM by moving the pinhole to all possible axial locations. Intensity diffracted through the same CPM from an object placed within the axial limits of the PSH library is recorded by a digital camera. The recorded intensity this time is composed as the object hologram. The image of the object at any axial plane is reconstructed by cross-correlating the object hologram with the corresponding component of the PSH library. The reconstruction noise attached to the image is suppressed by various methods. The reconstruction results of multiplane and thick objects by this technique are compared with regular lens-based imaging.

## Introduction

Digital holography systems are found to possess various advantages compared to their regular imaging counterparts[1-4], for instance, their inherent three-dimensional (3D) imaging capabilities. Contrary to regular imaging, digital holography is an indirect multi-step imaging process; recording the hologram of an object using an optical system in the first step, followed by digital processing and image reconstruction in the following steps[5,6]. This indirect procedure offers a platform for various opportunities of signal processing[7-10].

In the recent years, there has been an increasing interest among researchers in incoherent digital holography systems due to various benefits ranging from the use of low-cost light sources to enhancement in the image resolution[11-13]. One such well-established incoherent digital holography system called Fresnel incoherent correlation holography (FINCH)[14] makes use of the digital signal processing and self-referencing interference to achieve a lateral resolution beyond that of a conventional equivalent imaging system[15]. In FINCH, the light beam diffracted by the object is split into two; one beam passes through a quadratic phase mask while the other beam remains unmodulated and the two mutually coherent beams are interfered to create the hologram. The image is then digitally reconstructed by imitating the well-known optical propagation operators. The lateral resolution of optimal FINCH[15] is 2 times and 1.5 times higher than that of equivalent coherent and incoherent imagers, respectively. However, the axial resolution of FINCH is lower.

Another self-referenced interference based incoherent digital holography system called coded aperture correlation holography (COACH) was developed to increase the axial resolution at the expense of some loss of the lateral resolution[16]. In COACH, the quadratic phase mask of FINCH is replaced by a random-like coded phase mask (CPM). The COACH lateral and axial resolutions are equivalent to that of a regular imaging system with the same numerical aperture (*NA*). However, unlike regular imagers, COACH system can grab the entire 3D visual information in three camera shots and the entire 3D data is compressed to a 2D digital hologram, which enables one to store, transfer, and process the data more easily. In COACH, two holograms are synthesized, one is created once in the training stage of the system for a point object and is called point spread hologram (PSH). The other, for an object, is termed object hologram whereas both the point and the object are located at the same axial location and recorded with identical optical conditions. The object is reconstructed by a cross-correlation of the object hologram and the PSH. Therefore, for reconstruction of a multiplane, or 3D object, a library of PSHs are prerecorded at all possible axial locations and the visual information of the object at any plane can be accurately reconstructed by correlating the corresponding PSH with the object hologram. Lately, a modified COACH system for 3D imaging and simultaneous wavelength sensing was developed by amplifying the wavelength sensitivity with a diffractive objective lens[17]. In this case, the PSHs were cataloged for different axial locations and also for different wavelengths.

In this line of research, an interferenceless COACH (I-COACH) system was developed recently for 3D imaging without two wave interference[18]. This advancement has opened up many possibilities in the field of holography by defying one of the fundamental requirements of holography which is the use of interferometers. In I-COACH, the 3D visual information of the object is stored and retrieved without two-wave interference. Moreover, the interferenceless scenario simplifies the optical configuration, as cumbersome requirements are not needed to be satisfied. There is no need to isolate the system from external vibrations, it is not essential to align the optical system for precise overlap between the interfering beams, and the optical power efficiency becomes higher when the input beam is not split[19,20].

In this study, a further advancement in the interferenceless incoherent digital holography is proposed by demonstrating a lensless I-COACH (LI-COACH). In other words, we present a new lensless, interferenceless, incoherent digital holography system for 3D imaging. Lensless imaging systems can offer aberration free imaging and, in particular, to solve problems associated with the manufacturing of objective lenses for telescopes and microscopes.[21,22] This technique can be a benefit for telescopic and microscopy systems for visible as well as non-visible wavelengths for which fabrication of lenses and creating interference are challenging tasks. Lensless and interferenceless imaging systems have been already proposed by Chi et.al.[23,24] However, digital holograms of a 3D scene have not been recorded in their cases and their experimental results have had a high level of background noise because of lack of any noise reduction mechanism. Other closely related works propose various methods of the point spread function engineering for 3D imaging[25-32]. However, these works cannot record digital holograms of the 3D scene as LI-COACH offers. Lensless incoherent digital holography systems[33,34] with wave interferences developed in the past and the present study shows that lensless holographic 3D imaging can be done without two-wave interference. The optical configuration of LI-COACH is simple as a regular imaging system but with the 3D imaging capability. As in the previous cases[18], the 3D visual information of the scene is also compressed into a single 2D hologram. However, unlike the previous cases[18], the present hologram is synthesized from two, rather than three, camera shots. Another advantage of the LI-COACH system is the increased field of view (in comparison to[18]) achieved because there is only a single optical component between the object and the digital camera.

## Methodology

The optical configuration of LI-COACH is shown in Fig. 1. The light from an incoherent light source critically illuminates an object using a refractive lens $L$. Obviously, this illumination setup with the lens is not part of LI-COACH. The object can be illuminated by different illumination systems or can be a self-luminous object. The only condition which should be constrained is a complete spatial incoherence in the object space. The light diffracted by the object is modulated by a random-like CPM and reaches the image sensor on which the detected intensity pattern is recorded. The CPM in this case is designed using modified Gerchberg-Saxton algorithm (GSA)[35] with Fresnel propagators instead of Fourier transforms like in the original GSA. The aim of the GSA is to reduce the background noise during the hologram reconstruction. Fig. 2 shows the block diagram of the GSA to generate the CPM. In the beginning, a random phase $\Phi(x,y)$ is generated and multiplied with the diverging spherical wave (in the paraxial approximation) originated from a point object located on the optical axis at a distance $z_s$ from the CPM. After propagating from the CPM to the sensor plane using the Fresnel free-space propagator, the phase distribution of the obtained complex amplitude is extracted and the magnitude is replaced by a zero-padded uniform matrix. The zero-padded uniform matrix has been used to constrict the field diffracted from the CPM to have a uniform intensity over a limited area at the center of the sensor plane. The resulting complex amplitude is back propagated to the CPM plane using the Fresnel back propagator. The phase of the complex amplitude is extracted and attached to a constant magnitude, for the next iteration. This cycle is repeated until the generated intensity profile at the sensor plane converges towards the constraint. The obtained phase from the GSA is multiplied with $Q(-1/z_s)$ in order to cancel the effect of input spherical wave phase $Q(1/z_s)$. In the current setup, 20 iterations were found to be sufficient as the decrease in the background noise value is negligible beyond the 20 iterations.

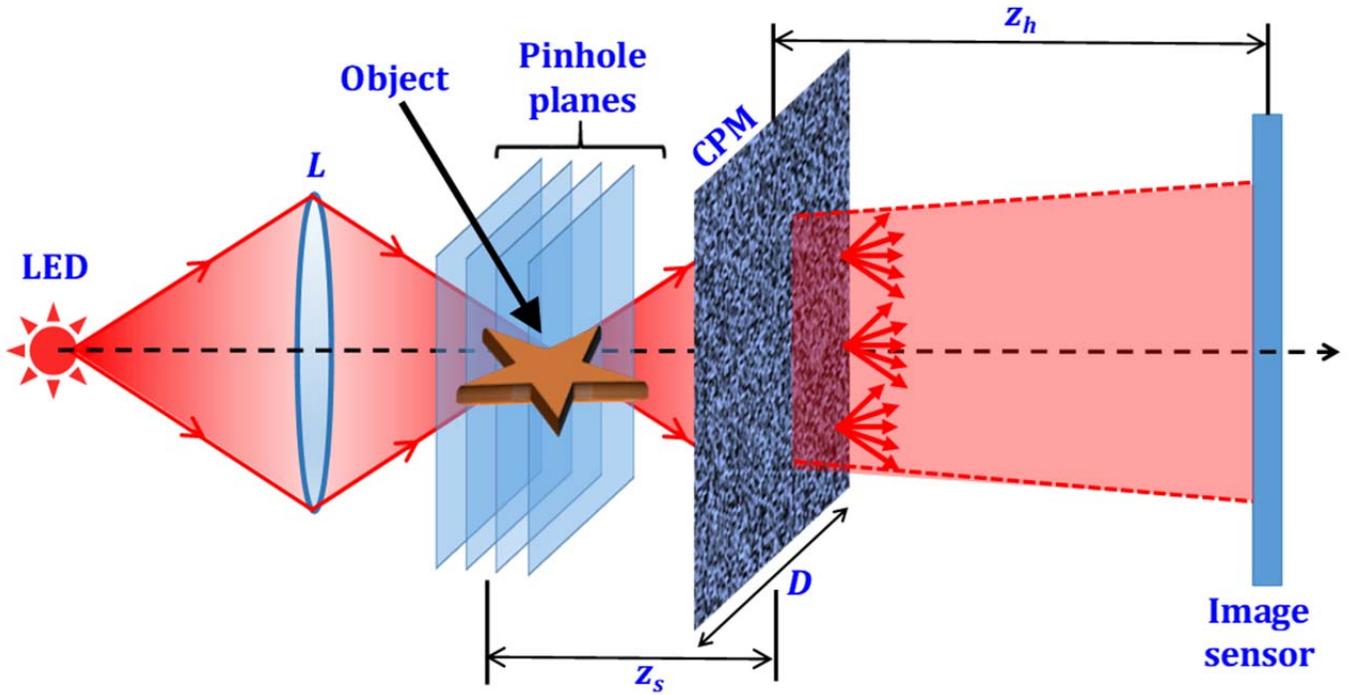

Fig. 1. Optical configuration of LI-COACH.

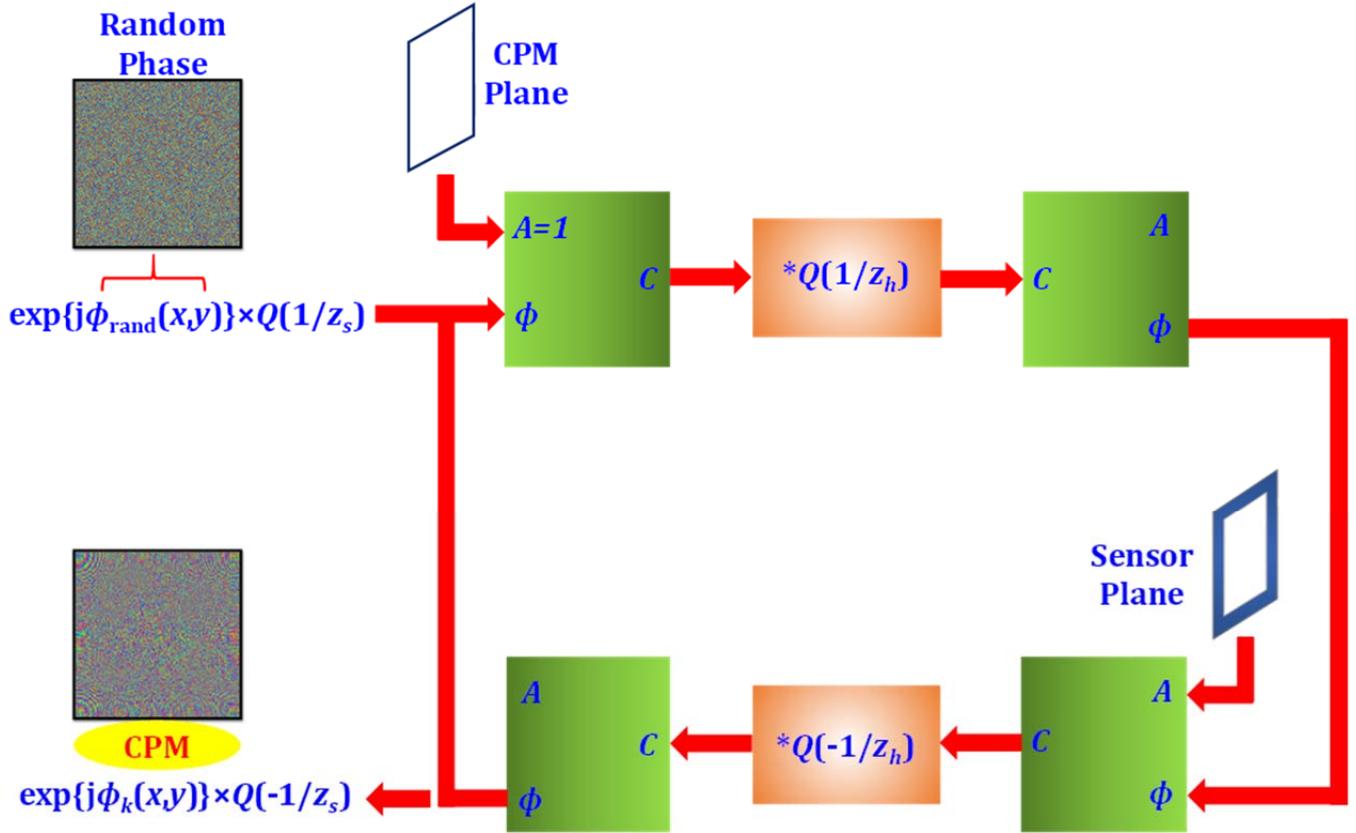

Fig. 2. Block diagram of GSA to calculate the CPM.

In order to improve the signal to noise ratio (SNR) in the reconstruction plane, the autocorrelation of the PSH should be as close as possible to δ function. This is because the autocorrelation of the PSH is a reconstruction of a point object and a reconstruction of an arbitrary object is actually a collection of PSH autocorrelations shifted according to the locations of the object points. In order to obtain the PSH autocorrelation closest to δ function, the magnitude of the PSH spectrum should be as close as possible to a different-than-zero constant. However, any single intensity pattern cannot have a spectrum with a constant magnitude because of the intense peak in the center of the spectrum which is higher than all its surroundings. Therefore, at least two independent intensity patterns should be recorded corresponding to two different CPMs generated from two different initial independent random phase masks. The Fourier transform of the subtraction between the intensity patterns can possess approximately a constant magnitude forced by the GSA. However, the present GSA is limited in achieving the ideal δ-like PSH autocorrelation, and the reconstructed image contains background noise due to the GSA limitations[35]. The phase-only filtering (POF) correlation technique and averaging technique were implemented in order to reduce the background noise further and to improve the SNR in the reconstructed image[36]. In the POF technique, the spatial spectrum of the object hologram is filtered with the POF. In the averaging technique, a collection of PSHs and object holograms using different CPMs are recorded. The resulting multiple complex reconstructed images are averaged to produce an image with an improved SNR[17,36].

The following analysis is based on the configuration of Fig. 1. The light diffracted from a point object with an amplitude of $\sqrt{I_s}$ located at $(\bar{r}_s, z_s) = (x_s, y_s, z_s)$ reaches the CPM with a complex amplitude $\sqrt{I_s}$ $C_1 L(\bar{r}_s/z_s) Q(1/z_s)$, where $C_1$ is a complex constant, $L$ and $Q$ represent linear and quadratic phase functions, given by $L(\bar{s}/z) = \exp\left[i 2\pi (\lambda z)^{-1} (s_x x + s_y y)\right]$ and $Q(a) = \exp\left[i\pi a \lambda^{-1} (x^2 + y^2)\right]$, respectively. The complex amplitude after the CPM is given by $\sqrt{I_s} C_1 L(\bar{r}_s/z_s) \cdot Q(1/z_s) \cdot \exp\left[i\Phi_k(\bar{r})\right]$, where $\Phi_k(\bar{r})$ is the $k$-th quasi-random phase CPM calculated using the modified GSA shown in Fig. 2 and $k=1,2$. The complex amplitude at the image sensor is given by a convolution of $\sqrt{I_s} C_1 L(\bar{r}_s/z_s) \cdot Q(1/z_s) \cdot \exp\left[i\Phi_k(\bar{r})\right]$ with $Q(1/z_h)$. Therefore, the $k$-th intensity pattern on the image sensor is given by,

$$I_k(\bar{r}_0; \bar{r}_s, z_s) = \left| \sqrt{I_s} C_1 L\left(\frac{\bar{r}_s}{z_s}\right) Q\left(\frac{1}{z_s}\right) \exp\left[i\Phi_k(\bar{r})\right] * Q\left(\frac{1}{z_h}\right) \right|^2$$
$$= I_k\left(\bar{r}_0 - \frac{z_h}{z_s} \bar{r}_s; 0, z_s\right)$$
(1)

where the asterisk sign denotes a two-dimensional convolution and $\bar{r}_0 = (u, v)$ is the transverse location vector on the sensor plane. The second equality of Eq. (1) indicates that the intensity on the sensor plane is a shifted version of the intensity response for a point object located on the optical axis $(\bar{r}_s = 0)$, where the distance of the shift is $\bar{r}_s z_h/z_s$.

A 2D object in the object plane can be considered as a collection of $N$ object points given by

$$o(\bar{r}_s) = \sum_j^N a_j \delta(\bar{r} - \bar{r}_{s,j})$$
(2)

When the object is illuminated by an incoherent quasi-monochromatic light source, there is no interference between the various responses of the form given by Eq. (1). Hence, the overall intensity distribution on the sensor plane is a sum of the responses of the individual points, given by,

$$I_{OBJ,k}(\bar{r}_0; z_s) = \sum_j a_j I_k\left(\bar{r}_0 - \frac{z_h}{z_s}\bar{r}_{s,j}; 0, z_s\right) \quad (3)$$

Both $I_{OBJ,k}(\bar{r}_0; z_s)$ and $I_k(\bar{r}_0; z_s)$ are real positive quasi-random functions. The recovery of the image can be performed by a cross-correlation between $I_k(\bar{r}_0; z_s)$, the intensity response to a point, and $I_{OBJ,k}(\bar{r}_0; z_s)$, the intensity response to the object. However, a cross-correlation between two real positive yields undesired background distribution on the recovered image. The condition that can minimize the background is that the autocorrelation of the PSH should be as close as possible to the $\delta$ function. From the convolution theorem, this condition can be satisfied if the magnitude of the Fourier transform the PSH is uniform and equal to some constant greater than zero. This property cannot be achieved with real positive PSH because of the value at the origin of the spectrum which is much intense than any other value. Only a superposition of $K \geq 2$ intensity responses can approximately satisfy the constraint of the uniform spectrum. In the previous case[18], a superposition of three ($K=3$) intensity responses was used. Here we show that $K=2$ is enough to yield acceptable results, with relatively high SNR, if one intensity is subtracted from the other. This result is obtained because the averages of both intensities are approximately the same and hence the subtraction minimizes the average of the PSH such that the magnitude of the PSH spectrum can be closer to a uniform value. Therefore, in order to minimize this background distribution both $H_{PSH}(\bar{r}_0; z_s)$ and $H_{OBJ}(\bar{r}_0; z_s)$ the PSH and the object holograms, respectively, are calculated as follows,

$$H_{PSH}(\bar{r}_0; z_s) = I_1(\bar{r}_0; z_s) - I_2(\bar{r}_0; z_s) \quad (4)$$

Similarly,

$$\begin{aligned} H_{OBJ}(\bar{r}_0; z_s) &= I_{OBJ,1}(\bar{r}_0; z_s) - I_{OBJ,2}(\bar{r}_0; z_s) \\ &= \sum_j a_j I_1\left(\bar{r}_0 - \frac{z_h}{z_s}\bar{r}_{s,j}; 0, z_s\right) - \sum_j a_j I_2\left(\bar{r}_0 - \frac{z_h}{z_s}\bar{r}_{s,j}; 0, z_s\right) \\ &= \sum_j a_j H_{PSH}\left(\bar{r}_0 - \frac{z_h}{z_s}\bar{r}_{s,j}; z_s\right), \end{aligned} \quad (5)$$

The background noise can be minimized if the autocorrelation of PSH yields a function close to a delta function with minimum side lobes. This condition is achieved using the modified GSA by having a uniform intensity on the CPM plane and on the sensor plane inside a predefined area. As mentioned above, the spectral constraint of the PSH can be satisfied by choosing $K=2$ and subtracting one intensity pattern from the other. Therefore, two intensity patterns are recorded for both the object and the point object using two CPMs calculated with different initial random phases.

The image is reconstructed by correlating $H_{OBJ}(\bar{r}_0;z_s)$ with $H_{PSH}(\bar{r}_0;z_s)$ as follows,

$$\begin{aligned} P(\bar{r}_R) &= \iint H_{OBJ}(\bar{r}_0;z_s) H^*_{PSH}(\bar{r}_0-\bar{r}_R;z_s) d\bar{r}_0 \\ &= \iint \sum_j a_j H_{PSH}\left(\bar{r}_0 - \frac{z_h}{z_s}\bar{r}_{s,j};z_s\right) H^*_{PSH}(\bar{r}_0-\bar{r}_R;z_s) d\bar{r}_0 \\ &= \sum_j a_j \Lambda\left(\bar{r}_R - \frac{z_h}{z_s}\bar{r}_{s,j}\right) \approx o\left(\frac{\bar{r}_s}{M_T}\right). \end{aligned} \qquad (6)$$

where $\Lambda$ is a $\delta$-like function, $\sim 1$ at (0,0) and $\sim 0$ elsewhere. It must be noted that the image is reconstructed using the cross-correlation [Eq. (6)], and therefore, the transverse and axial resolutions are dictated by the transverse and axial correlation lengths, determined by the width and the length of the smallest spot that can be recorded on the sensor plane by the spatial light modulator (SLM) with an active area of diameter of $D$. Therefore, the transverse and axial resolutions are approximately $1.22\lambda z_s/D$ and $8\lambda(z_s/D)^2$ respectively, which is the same as that of regular imaging with the same *NA*. The magnification of the imaging system implied from Eq. (6) in $M_T=z_h/z_s$.

## Experiments

The LI-COACH system with two illumination channels is experimentally demonstrated using the digital holography setup shown in Fig. 3. The experimental setup consists of two light emitting diodes (LEDs) (Thorlabs LED631E, 4mW, $\lambda$ = 635 nm, $\Delta\lambda$ = 10 nm) mounted on two illumination channels. Two identical lenses, L1 and L2, were used to illuminate the objects in channel 1 and channel 2, respectively. The distance between the lenses L1 and L2 and the respective objects is 18 cm. Two negative National Bureau of Standards (NBS) (NBS 1963A Thorlabs) resolution charts (RCs) were used as objects and are mounted in channel 1 and 2. Element 5.6 *lp/mm* of NBS1 RC mounted in channel 1 and elements 6.3 and 7.1 *lp/mm* of NBS2 RC mounted in channel 2 were critically illuminated. The light diffracted by the two objects (NBS1 and NBS2 RCs) were combined using a beam splitter BS1. The combined beams pass through the polarizer P, oriented along the active axis of SLM (Holoeye PLUTO, 1920×1080 pixels, 8 μm pixel pitch, phase-only modulation) in order to enable full modulation of the light. An iris with a radius of 0.8 cm was mounted just before the SLM to control the size of the NA. The NA of the illumination system is approximately 0.016 ($NA=D/2z_s$, where D is the diameter of the aperture D=0.8 cm; $z_s$=26 cm). The minimum resolved element has the width of about 25 μm ($0.61\lambda/NA$) and the length of 5.4 mm [$2\lambda/(NA)^2$]. A pinhole with a diameter of ~100 μm was used as the point object in channel 1. The maximum lateral resolution (25 μm) achievable with the current optics configuration is sacrificed by using a larger pinhole (100 μm) in order to achieve a detectable optical power at the image sensor while recording the PSHs. Two different CPMs generated from two different initial random phase masks calculated by GSA were displayed on the SLM with 1080×1080 pixels and two intensity patterns corresponding to these two CPMs were recorded by CMOS camera [pco.edge 5.5 scientific CMOS (sCMOS), 2560×2160 pixels, 6.5 μm pixel pitch]. The distance between the objects (NBS1 and NBS2 RCs) and the SLM was 26 cm. The distance between the SLM and the camera was $z_h$=26 cm.

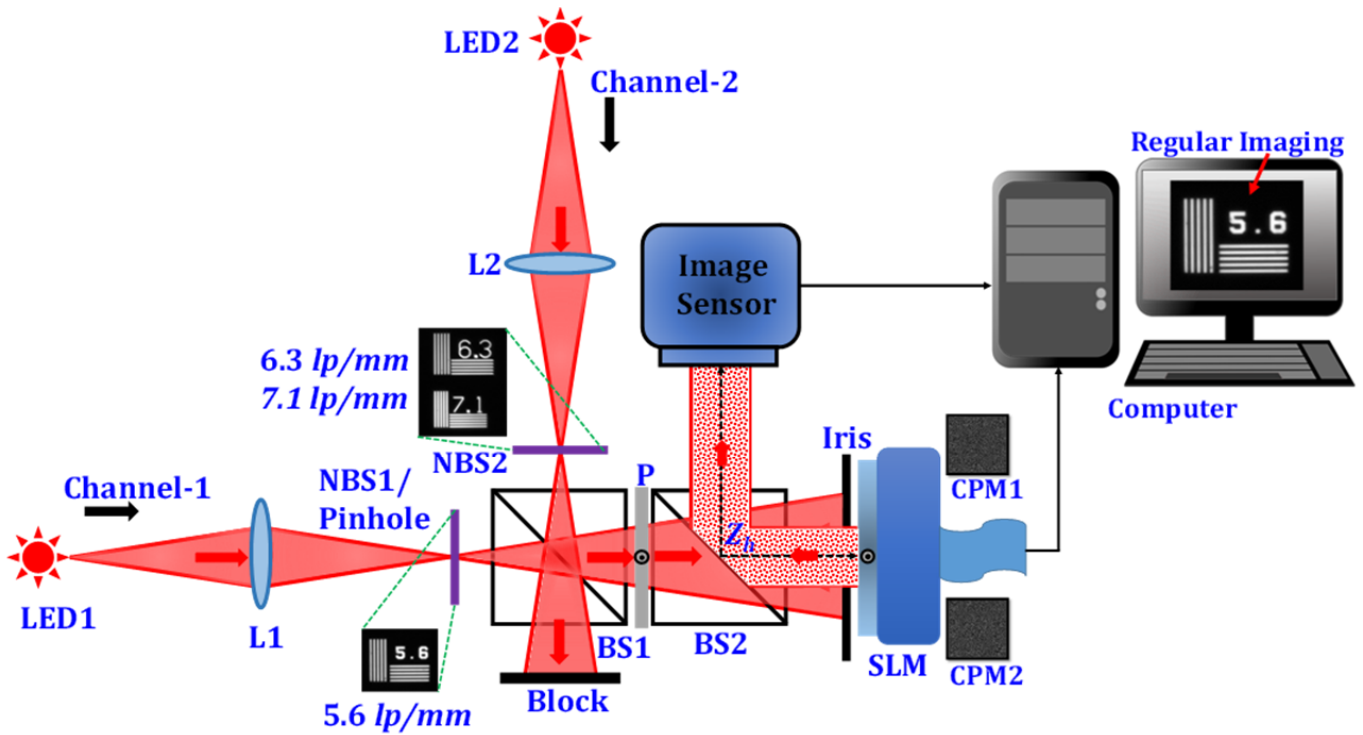

Fig. 3. Experimental setup of LI-COACH with two illumination channels.

## Experimental Results

To investigate LI-COACH, various experiments were carried out with the optical setup of Fig. 3. In the first experiment, 40 different CPMs were synthesized and the corresponding intensity patterns were recorded using the pinhole mounted at the back focal plane of the lens L1 in channel 1 and blocking the channel 2. From the 40 intensity recordings, 20 PSHs were synthesized by subtracting every two intensity patterns. Similarly, 40 intensity patterns were recorded by replacing the pinhole by the NBS object (5.6 $lp/mm$) and 20 object holograms were synthesized using the intensity patterns from the same CPM pairs as used for the PSHs. The image of the object is reconstructed by a cross-correlation of the object hologram and the PSH. The complex distributions of the different reconstructions were averaged to minimize the background noise.

Phase images of one pair of CPMs, corresponding to $k=1$ and $k=2$, are given in Fig. 4(a) and 4(b). Two intensity patterns of the pinhole ($I_{PSH1}$ and $I_{PSH2}$) and two object intensity patterns ($I_{Object1}$ and $I_{Object2}$) of the element 5.6 $lp/mm$ recorded with the same CPM pair are shown in Figs. 4(c)-4(d), and Figs. 4(e)-4(f), respectively. The two intensity patterns $I_{PSH1}$ and $I_{PSH2}$ were subtracted one from the other to obtain a point spread hologram ($H_{PSH}$). Similarly, an object hologram ($H_{OBJ}$) is obtained by subtracting two object intensity patterns ($I_{Object1}$ and $I_{Object2}$). The images of the PSH and the object hologram are shown in Figs. 4(g) and 4(h), respectively. The reconstructed image of the NBS object from a single hologram is shown in Fig. 4(i). This result was obtained by a cross-correlation of the object hologram and the POF version of the PSH, both are synthesized from the same CPM pair. Finally, the 20 reconstructed images were averaged and the resulting reconstructed image is shown in Fig. 4 (j). Fig. 4(k) shows the regular imaging for the element 5.6 $lp/mm$ of the NBS object. The average background noise on the reconstructed images for different numbers of CPMs were calculated and plotted in Fig. 5. It is observed that the background noise decreases over the reconstruction plane with the increase in the number of CPMs, and become almost constant for averaging over 20 CPM pairs.

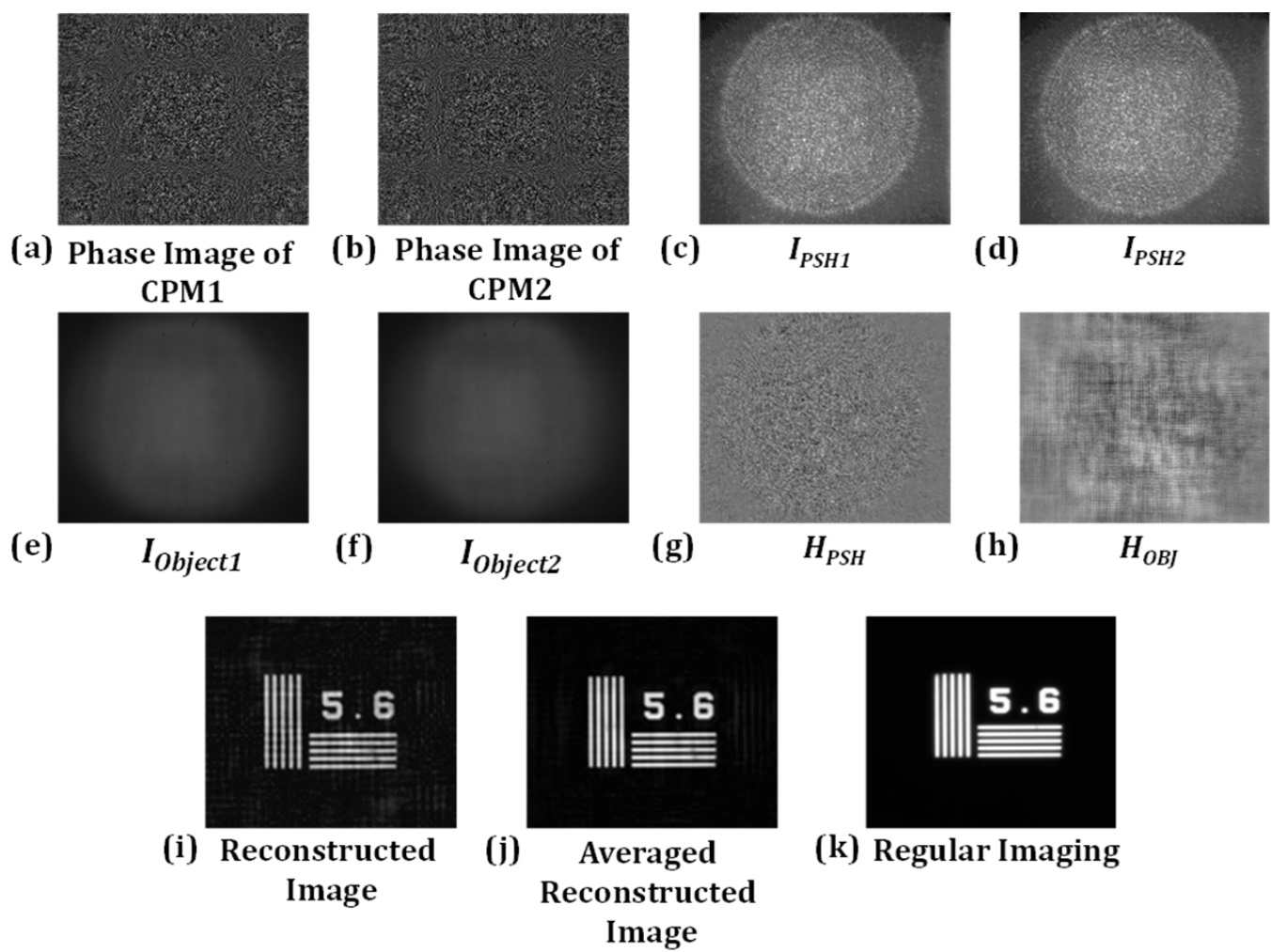

Fig. 4. (a)-(b) Phase images of the CPM1 and CPM2; (c)-(d) intensity patterns of PSHs corresponding to CPM1 and CPM2 respectively; (e)-(f) intensity patterns of object holograms corresponding to CPM1 and CPM2 respectively; (g)-(h) magnitude of the hologram of the pinhole and the NBS RC; (i) reconstructed image; (j) resultant image obtained after averaging over 20 LI-COACH reconstructed images; and (k) regular imaging.

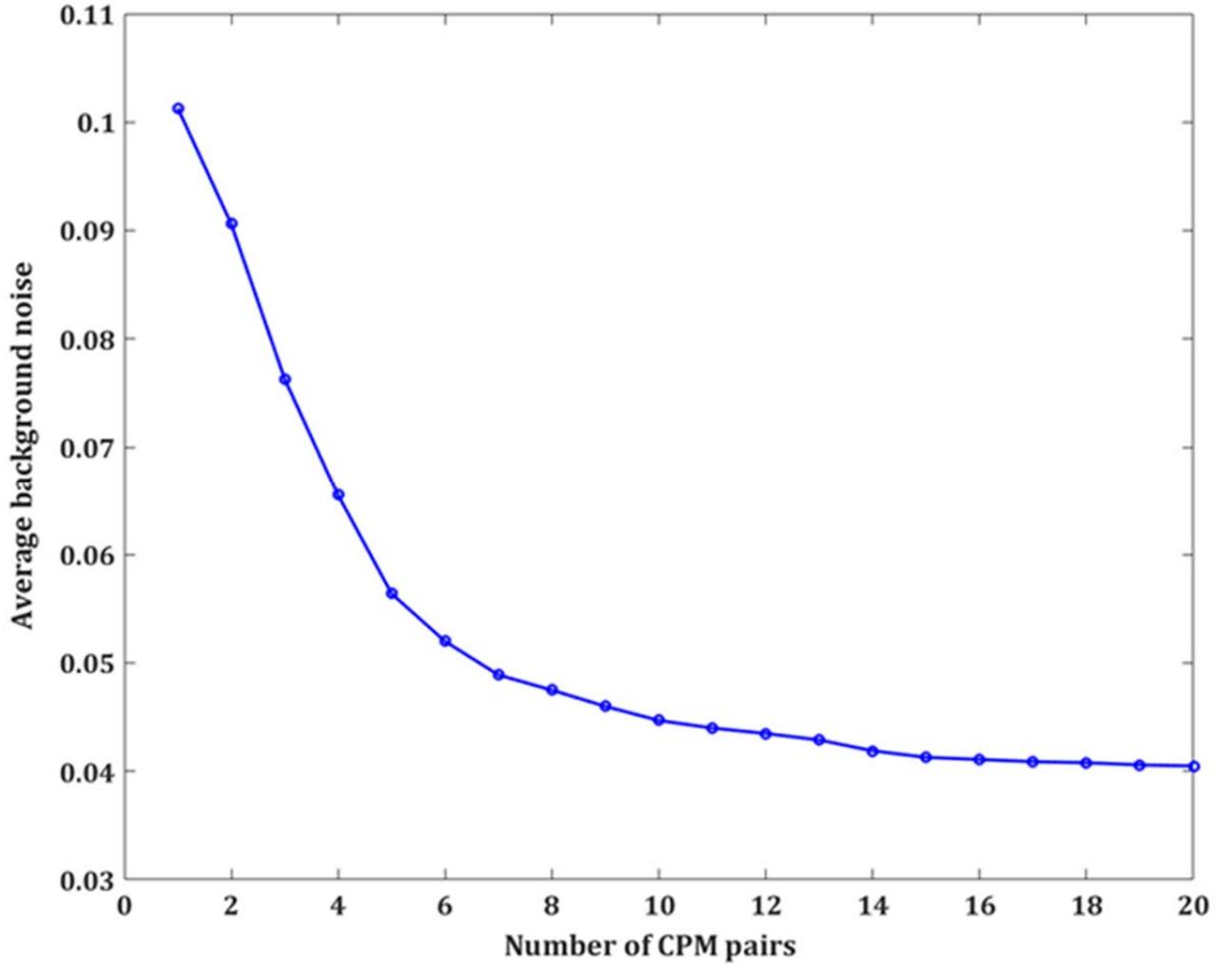

Fig. 5. Plot of average background noise on the reconstructed images averaged over different numbers of CPM pairs

In the next experiment, the axial and lateral distribution of imaged pinholes from the LI-COACH system and from the regular imaging system were measured and compared. The axial location of the pinhole was varied from -28 mm to +28 mm with respect to the back focal plane of lens L1 and the corresponding PSHs were recorded at every location of the pinhole. The reconstructed images were obtained by correlating the recorded PSHs with the PSH recorded for the pinhole at the back focal plane of the lens L1. The intensity of the reconstructed images at $(x,y)=(0,0)$ were measured and plotted against the axial locations of the pinhole. In the case of regular imaging, for every axial location of the pinhole, the image of the pinhole was recorded and the image intensity at $(x,y)=(0,0)$ was measured and plotted. Fig. 6 shows the plots of the intensity values of reconstruction/imaging at $(x,y)=(0,0)$ resulting from LI-COACH and regular imaging. The similarity between the two axial distributions indicates the resemblance in the axial resolution of the two imaging methods.

The lateral distribution of imaged points by the LI-COACH is measured and compared with the regular imaging by placing a two-point object at the back focal plane of the lens L1. To obtain the reconstructed image, the recorded hologram of the two-point object was correlated with the PSHs prerecorded with the pinhole at the back focal plane of the lens L1. The intensity of the reconstruction/imaging of the two-point object is plotted in Fig. 7.

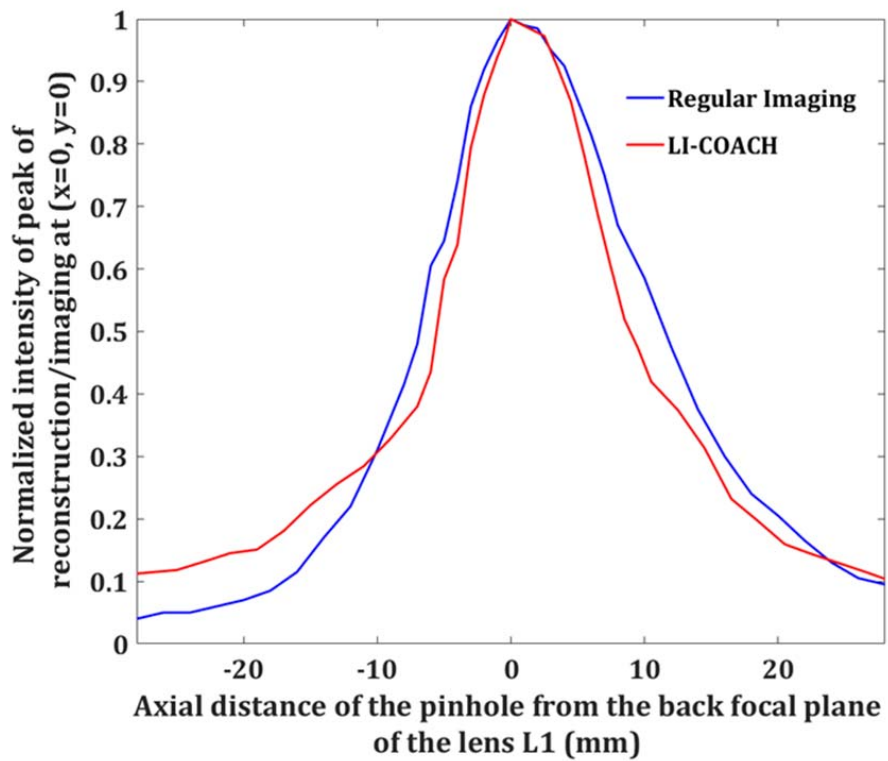

Fig. 6. Normalized intensity of reconstruction/imaging at ($x$=0, $y$=0) versus the axial distance of the pinhole from the back focal plane of the lens L1.

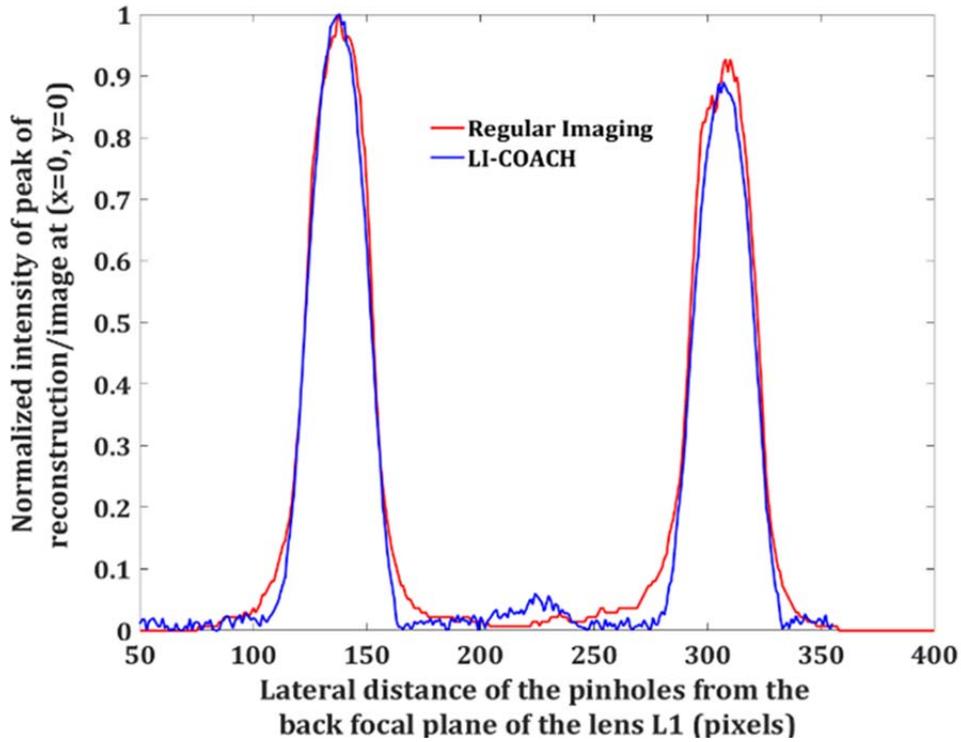

Fig. 7. Normalized intensity of reconstruction/imaging of two-point object along the lateral axis of the back focal plane of the lens L1.

Figures 7 indicates that the lateral resolution as well as the image magnification of LI-COACH, are about the same as that of the regular imaging system.

In the third experiment, a library of PSHs were created by moving the pinhole to 17 axial locations (-40 to +40 mm) in steps of 5 mm with respect to the back focal plane of the lens L1 in channel 1. Forty intensity patterns were recorded at every axial location corresponding to the 20 CPM pairs synthesized using the GSA and composed them into 20 PSHs for averaging. So the library consists of 340 elements i.e., 20 PSHs at every axial location for 17 axial planes. Two NBS RCs were mounted in the two channels and the object holograms were recorded. It should be noted that both the NBS RCs were aligned in such a way that the element 5.6 *lp*/*mm* of the NBS1 RC and elements 6.3 and 7.1 *lp*/*mm* of NBS2 RC can be imaged onto the camera without lateral overlap. Next, the axial location of NBS1 RC was varied from -40 to +40 mm, in steps of 5 mm, with respect to the back focal plane of the lens L1, while the axial location of the NBS2 was constant. Object intensity patterns of the two-plane object were recorded corresponding to the same CPMs used for recording PSHs. The different planes of the object are reconstructed by correlating the corresponding PSHs from the library. The regular imaging and reconstruction results of LI-COACH for the two-plane object after averaging over 20 independent reconstructions are shown in Fig. 8. The experimental results reveal that the performance of the proposed LI-COACH is similar to that of regular imaging.

In the last experiment, the imaging of 3D objects constructed from two identical LEDs (LED-A and LED-B) and two one-dime coins (coin 1 and coin 2) both are separated by an axial distance of 15 mm is carried out. The experimental setup was then modified such that the 3D object can be illuminated critically in reflection mode by the illumination system in channel 2 using a different biconvex lens of focal length of 88.30 mm in order to obtain a larger illumination area on the surface of the 3D object. Figure 9 shows the modified experimental setup for imaging the 3D object in reflection mode. Once again 40 intensity patterns of the 3D object were recorded and composed into 20 object holograms, corresponding to the same CPM library used for recording the PSHs. The correlation of the synthesized complex holograms with the PSH libraries prerecorded at the back focal plane of the lens L1 and at $z_s$=-15 mm, reconstructs the information in the respective object planes. Figure 10(a) shows the regular imaging of the 3D object in which LED-B is focused and LED-A is out-of-focus, whereas in Fig. 10(b) LED-A is focused and LED-B is out-of-focus. The reconstruction results of LI-COACH obtained by correlations with the corresponding prerecorded PSHs, and after averaging over 20 independent reconstructions, are shown in Fig. 10(c) and 10(d) respectively. Figure 10(e) shows the photograph of the two LEDs system separated axially by 15 mm. Similarly, figure 11 shows the reconstruction results of LI-COACH after averaging over 20 reconstructions, regular imaging and the photograph of two one-dime coins system.

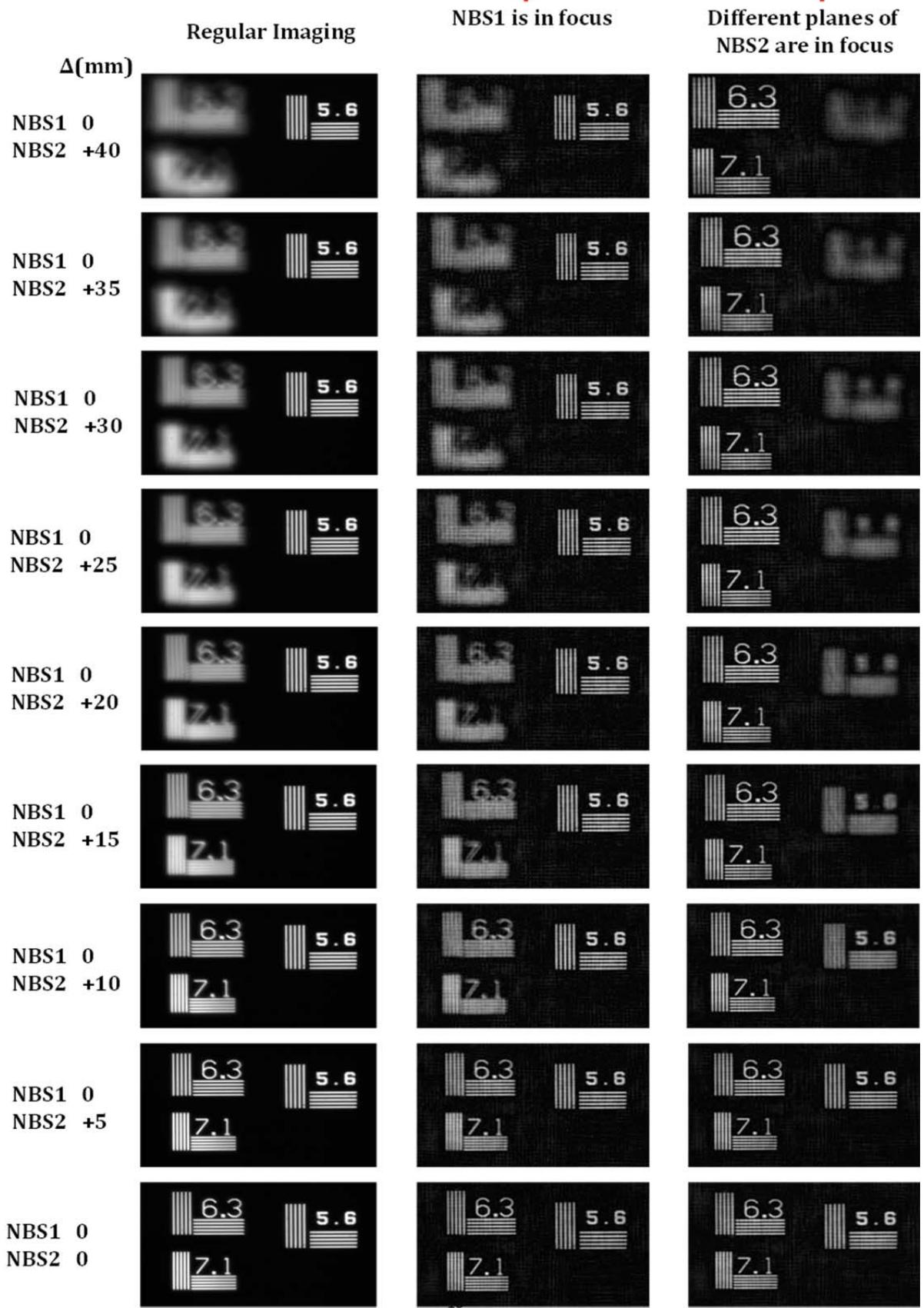

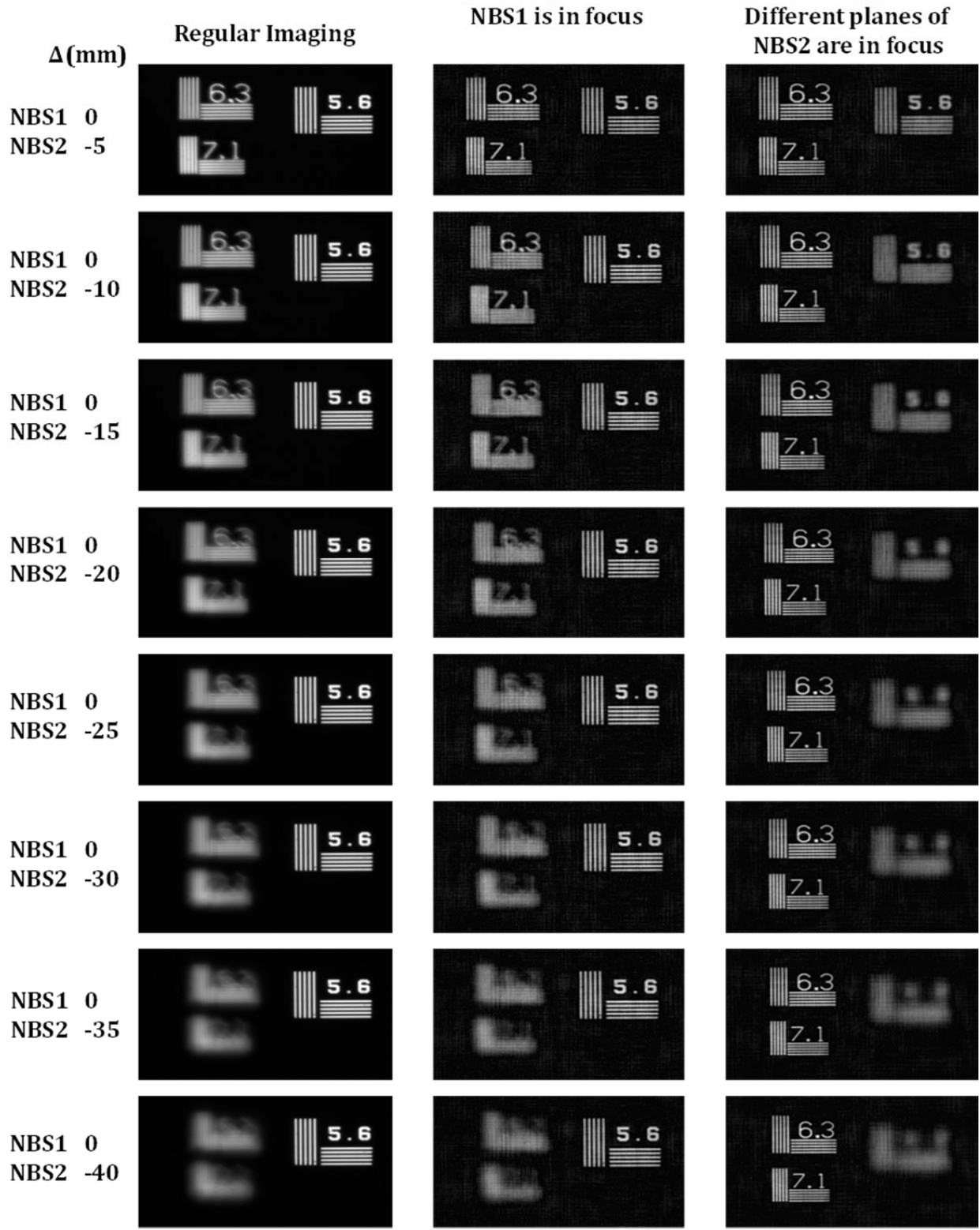

Fig. 8. Experimental comparison results of regular imaging and LI-COACH holograms recorded for a two-plane object made up of the NBS RCs by placing the NBS1 and NBS2 RCs in two lateral planes.

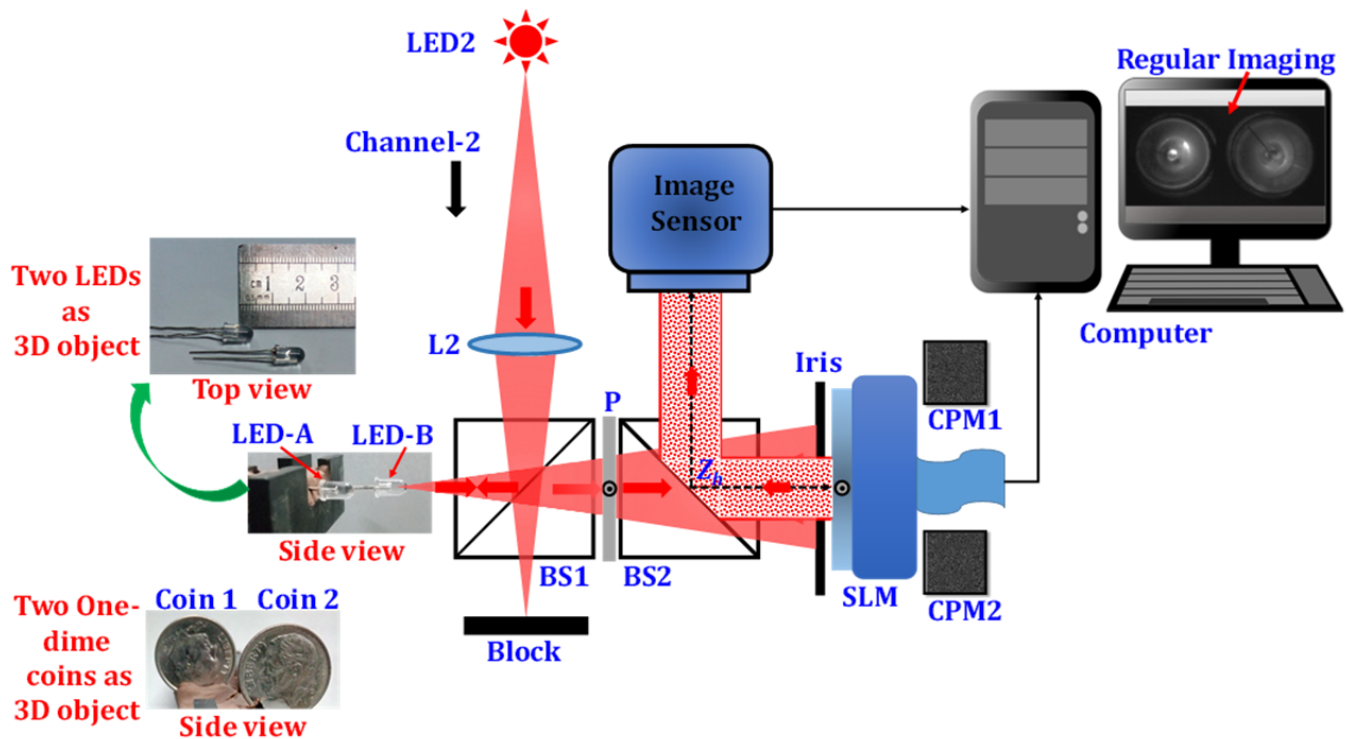

BS1, BS2 – Beam splitters; L2 – Refractive lens; LED – Light emitting diode; NBS – National bureau of standards; P – Polarizer; SLM – Spatial light modulator; ⊙ – Polarization orientation perpendicular to the plane of the page

Fig. 9. Experimental setup of LI-COACH for imaging reflective 3D object.

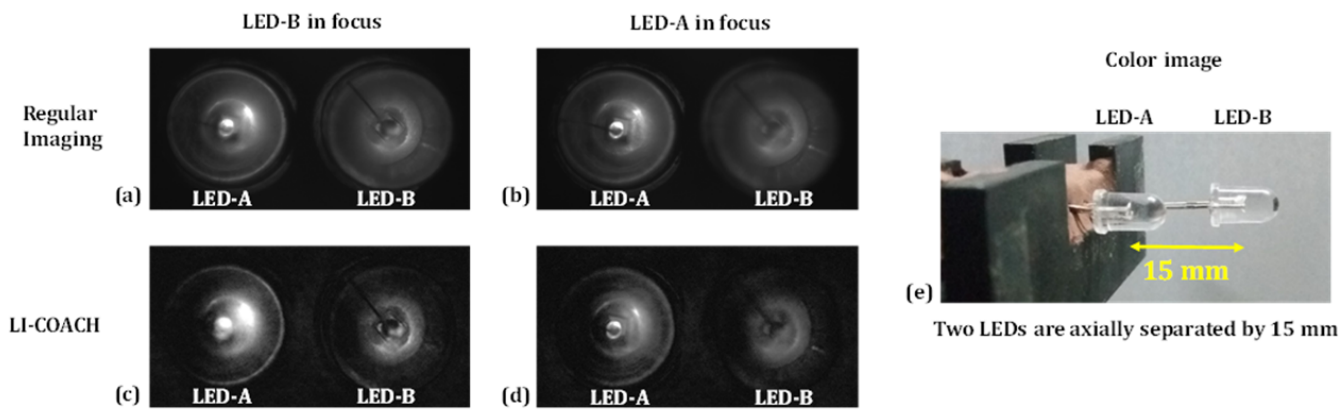

Fig. 10. Regular imaging and reconstruction results of LI-COACH of the different planes of the set of two LEDs separated by a distance of 15 mm.

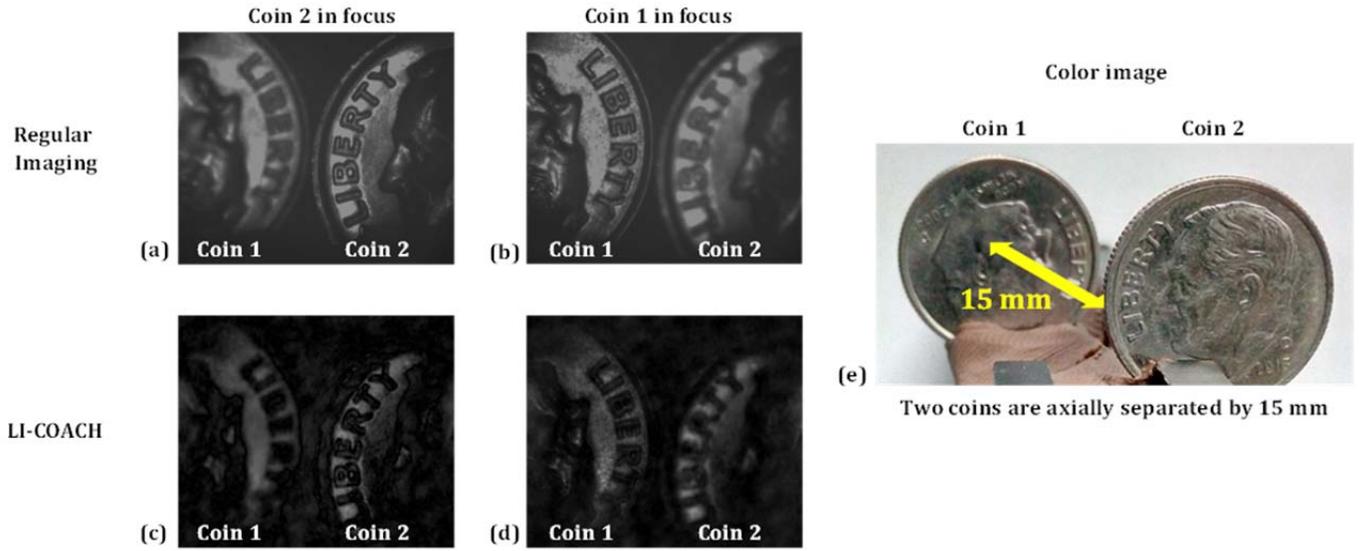

Fig. 11. Regular imaging and reconstruction results of LI-COACH of the different planes of the set of two one-dime coins separated by a distance of 15 mm.

## Conclusion

In conclusion, we have proposed and demonstrated a new inteferenceless and lensless incoherent digital holography system termed LI-COACH which provides 3D imaging without using any lens. The study demonstrates that in the case of indirect imaging one is not restricted to any conventional optical element like a lens or spherical mirror. A set of synthesized random-like phase masks can work as imaging element with reasonable SNR in a 3D imaging system.

A library of PSHs along some axial range is prerecorded using two synthesized CPMs. Then two intensity patterns resulting from the object after passing through the same CPMs, are recorded and composed to the object hologram. The visual information of the object at any plane can be obtained by cross-correlating the corresponding PSH and the object hologram. The noise attached to the reconstructed images was minimized using the POF and averaging technique. It must be noted that the PSH library needs to be created only once in the training stage of the system and can be used to reconstruct any number of images. The proposed system offers many advantages such as large field of view, simple, aberration-free, and compact setup. Moreover, the 3D imaging is achieved from a single point of view with only two camera shots and without any mechanical movement. In comparison to other self-interference holography systems such as FINCH and COACH, LI-COACH offers holograms acquisition without the complication of wave interference. The applicability of the idea was verified with two plane and thick objects and the preliminary results are promising. The LI-COACH technique can open up several possibilities for imaging applications.

## Acknowledgment

The work was supported by the Israel Science Foundation (ISF) (Grants No. 439/12, 1669/16).

## Author Contributions

J.R. carried out theoretical analysis and simulation for the research, M.K. and A.V. performed the experiments. The manuscript was written by M.K. and A.V. and J.R. All authors discussed the results and contributed to the manuscript.

## Additional Information

**Competing financial interests**: The authors declare no competing financial interests.